\documentclass[%
 	secnumarabic,
 	superscriptaddress,
 	numerical,
 	amssymb,
	frontmatterverbose, 
 	amsmath,
 	aps,
 	prl,
	floatfix,
    twocolumn,
]{revtex4-1}

\usepackage{ulem}
\usepackage{graphicx}
\usepackage[]{natbib}
\usepackage[usenames,dvipsnames]{color} 
\usepackage{dcolumn}
\usepackage{bm}
\usepackage{placeins}
\usepackage{epstopdf}
\usepackage{upgreek}
\usepackage{textcomp}
\usepackage{textgreek}
\usepackage{siunitx}
\usepackage[urlcolor=blue, hyperindex, colorlinks, bookmarks=true,linkcolor=black,citecolor=blue]{hyperref}

\newcommand{\boldit}[1]{\boldsymbol{#1}}

\begin{document}

\title[]{Mapping momentum-dependent electron-phonon coupling and non-equilibrium phonon dynamics with ultrafast electron diffuse scattering}

\author{Mark J. Stern}
 \thanks{These two authors contributed equally}
 \affiliation{
Department of Physics, Center for the Physics of Materials, McGill University, 3600 University Street, Montreal, QC, CA 
}%

\author{Laurent P. Ren\'{e} de Cotret} 
 \thanks{These two authors contributed equally}
 \affiliation{
Department of Physics, Center for the Physics of Materials, McGill University, 3600 University Street, Montreal, QC, CA 
}%

\author{Martin R. Otto} 
 \affiliation{
Department of Physics, Center for the Physics of Materials, McGill University, 3600 University Street, Montreal, QC, CA 
}%

\author{Robert P. Chatelain} 
 \affiliation{
Department of Physics, Center for the Physics of Materials, McGill University, 3600 University Street, Montreal, QC, CA 
}%

\author{Jean-Phillippe Boisvert} 
 \affiliation{
Department of Physics, Center for the Physics of Materials, McGill University, 3600 University Street, Montreal, QC, CA 
}%

\author{Mark Sutton} 
 \affiliation{
Department of Physics, Center for the Physics of Materials, McGill University, 3600 University Street, Montreal, QC, CA 
}%

\author{Bradley J. Siwick}
 \email{bradley.siwick@mcgill.ca}
 \affiliation{
Department of Physics, Center for the Physics of Materials, McGill University, 3600 University Street, Montreal, QC, CA 
}%
 \affiliation{
Department of Chemistry, McGill University, 801 Sherbrooke Street W, Montreal, QC, CA 
}%
\date{\today}


\begin{abstract}
Despite their fundamental role in determining material properties, detailed momentum-dependent information on the strength of electron-phonon and phonon-phonon coupling (EPC and PPC, respectively) across the entire Brillouin zone (BZ) has remained elusive~\cite{Devereaux2016}. Here we demonstrate that ultrafast electron diffuse scattering (UEDS) directly provides such information. By exploiting symmetry-based selection rules and time-resolution, scattering from different phonon branches can be distinguished even without energy resolution. Using graphite as a model system, we show that UEDS patterns map the relative EPC and PPC strength through their profound sensitivity to photoinduced changes in phonon populations. We measure strong EPC to the $K-$point transverse optical (TO) phonon of $A_1'$ symmetry ($K-A_1'$) and along the entire longitudinal optical (LO) branch between $\Gamma-K$, not only to the $\Gamma-E_{2g}$ phonon as previously emphasized~\cite{Pisana2007,Piscanec2004}. We also determine that the subsequent phonon relaxation pathway involves three stages; decay via several identifiable channels to transverse acoustic (TA) and longitudinal acoustic (LA) phonons ($1-2$~ps), intraband thermalization of the non-equilibrium TA/LA phonon populations ($30-40$~ps) and interband relaxation of the LA/TA modes (115~ps). Combining UEDS with ultrafast angle-resolved photoelectron spectroscopy (ARPES) will yield a complete picture of the dynamics within and between electron and phonon subsystems, helping to unravel complex phases in which the intertwined nature of these systems have a strong influence on emergent properties.
\end{abstract}

\maketitle

The nature of the couplings within and between lattice and charge degrees of freedom is a central concern of condensed matter and materials physics. Electron-phonon interactions play a dominant role in the electronic transport properties of metals~\cite{Grimvall1981}; they are the underlying cause of conventional superconductivity~\cite{BCS1957} and Peierls/Jahn-Teller instabilities~\cite{Peirels1930}. Furthermore, they are central to our understanding of the properties of many quasiparticles including polarons~\cite{Alexandrov2010} and phonon-polaritons~\cite{Dai2014}. Highly anisotropic (momentum-dependent) EPC has been identified as a key feature of superconductivity in MgB$_2$~\cite{Yildirim2001}. It is also intertwined with electron correlations in the iron-based superconductor FeSe~\cite{Gerber2017} and has been shown to contribute to the selection of the electronic ordering vector in some charge density wave materials including ErTe$_3$~\cite{Eiter2013} and NbSe$_2$~\cite{Zhu2015}. On the other hand, PPC dictates the thermalization properties of carrier/quasiparticle excitation energy. 

The subtle details of charge-lattice interactions can have an enormous impact on technologically-relevant material properties, but these interactions have so far resisted a comprehensive experimental investigation. Conventional ARPES~\cite{Tanaka2013}, inelastic x-ray/neutron scattering~\cite{Devereaux2016} and Raman spectroscopy~\cite{Yan2009} provide indirect information on the EPC strength through the shifting and broadening of spectral features only over a limited part of the Brillouin zone (BZ)~\cite{Devereaux2016}. Detailed studies of phonon-phonon interactions and decay in materials has typically been the province of theory~\cite{Bonini2007} or molecular dynamics simulations due to a lack of techniques capable of probing these interactions in any substantial detail. Time-domain approaches have recently opened new windows on the nature of EPC and PPC in materials. Time-resolved Raman spectroscopy has been used to directly measure the rate of energy exchange between photo-generated carriers and zone-center optical phonons and the subsequent relaxation of those specific non-equilibrium phonons~\cite{Tsen2009,Yan2009,Yang2017}. Time-resolved inelastic x-ray scattering at synchrotron and x-ray free electron laser facilities has provided a view of non-equilibrium distributions of off zone-center phonons in InP and InSb~\cite{Trigo2010} and the phonon band structure in Ge~\cite{Trigo2013,Zhu2015} through incoherent and coherent time-resolved diffuse scattering signals respectively.

\begin{figure}[t]
\centering
\includegraphics[width =\columnwidth]{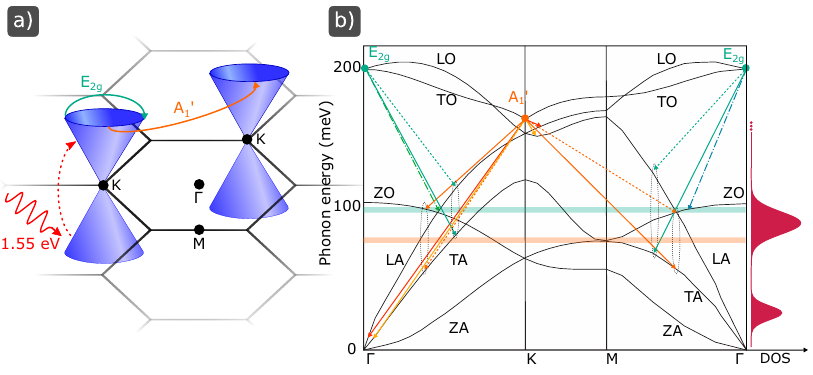}
\caption{EPC and PPC in graphite a) Simplified schematic of the electronic band structure of graphite illustrating the influence of the Dirac cones on the electron-phonon scattering process. In these experiments a pump laser pulse drives vertical electronic transitions ($\pi-\pi^{\star}$). The resulting hot electrons (see text) may inelastically scatter with the $\Gamma-E_{2g}$ phonon across a Dirac cone and with the $K-A_1'$ phonon between Dirac cones, conserving energy and momentum b) Phonon dispersion of graphite with the strongly coupled optical modes indicated; $\Gamma-E_{2g}$ (blue) and $K-A_1'$ (red). The dominant energy and momentum-conserving decay pathways are indicated with colored arrows. Dotted arrows should be thought of as going in the opposite momentum direction.  Side bar (red) provides a schematic of the non-equilibrium LA/TA phonon distribution produced through the decay of $\Gamma-E_{2g}$ and $K-A_1'$ phonons as determined by the UEDS data described in the text.}\label{fig:introfig}
\end{figure}

Here we demonstrate that ultrafast electron diffuse scattering (UEDS) using radio-frequency compressed electron pulses provides a general, lab-scale, time-resolved analog of diffuse x-ray scattering~\cite{Holt1999DeterminationSilicon}. This new method~\cite{Chase2016} is capable of directly determining both the relative momentum-dependent interaction strength between photogenerated carriers and phonons \textit{and} the subsequent phonon-phonon interactions governing the relaxation and thermalization of the excitation energy across the entire BZ with $\sim100~$fs time-resolution~\cite{Otto2017}. In pump-probe geometry~\cite{Chatelain2014a,Morrison2014a} these experiments map the transient changes to diffuse (inelastic) electron scattering patterns, which are themselves determined by the evolution of the non-equilibrium phonon distributions that follow electronic excitation.  We show that this technique is particularly well suited as a probe of 2D materials using thin graphite as a model system. Specifically, optical excitation at 800~nm with 35~fs laser pulses drives vertical electronic transitions of a $\pi-\pi^{\star}$ character on the well-known Dirac cones~\cite{CastroNeto2009} of single crystal graphite samples (Fig.~\ref{fig:introfig}~a). This excitation impulsively photo-dopes the material with a non-equilibrium electron-hole plasma with carrier density controllable by excitation fluence. TR-ARPES experiments have found that the first stage of relaxation is for the non-equilibrium distribution of carriers to thermalize internally through carrier-carrier scattering, forming a Fermi-Dirac distribution with well defined electron temperature within $\sim$50 fs~\cite{Gierz2015,Stange2015,Yang2017}. In this work we use UEDS patterns to determine how the energy stored in the hot electron system couples to the phonons and how the phonon system subsequently thermalizes, comparing the results\textemdash where possible\textemdash with earlier investigations using time-resolved Raman spectroscopy~\cite{Yan2009,Yang2017}, pump-probe spectroscopy~\cite{Ishioka2008,Kampfrath2005} and theory~\cite{Bonini2007}.

\begin{figure*}
\includegraphics[width=0.8\textwidth]{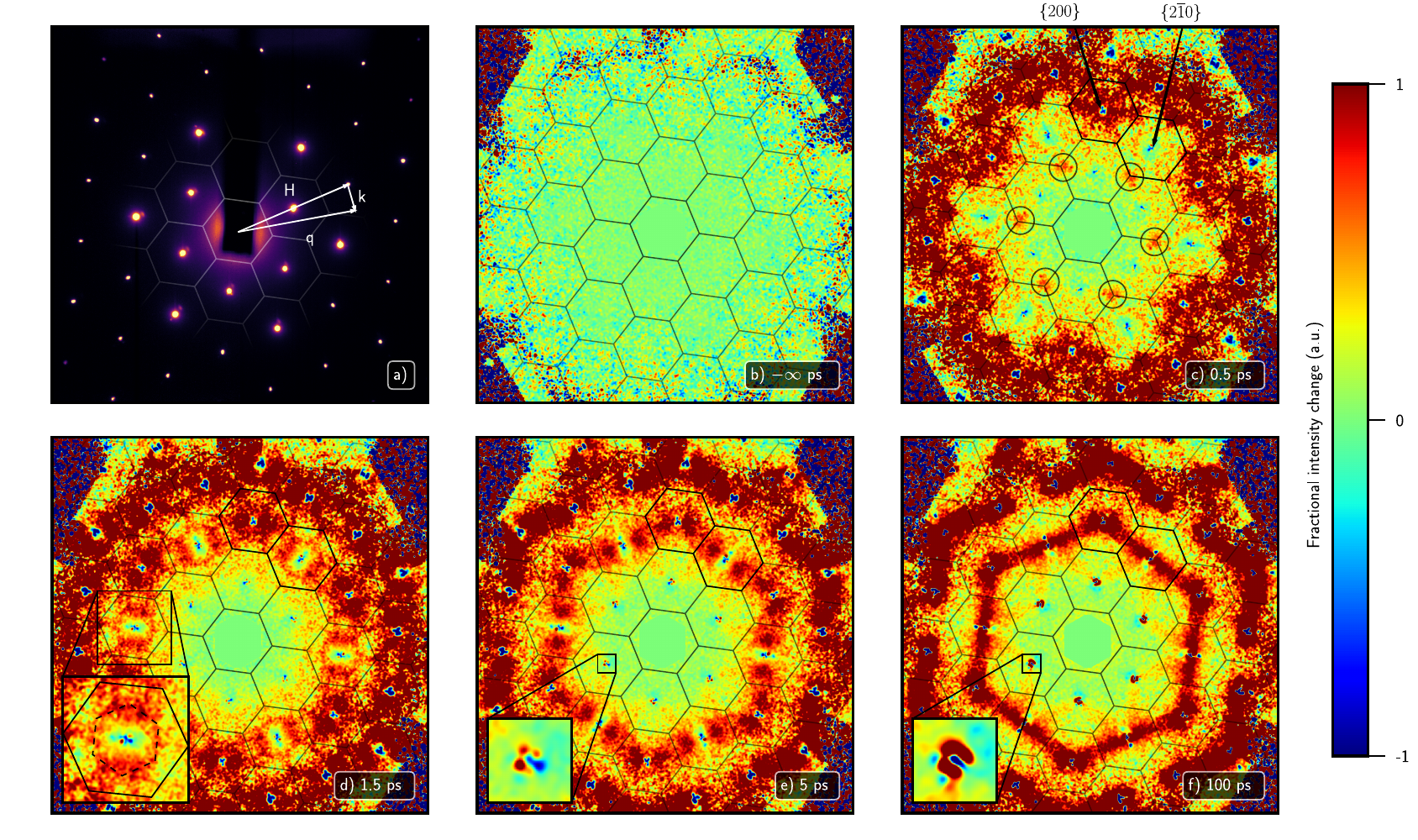}
\caption{Evolution of $\Delta I(\boldit{q},\tau)$ following photo-excitation of graphite (35~fs, 800~nm, 12~mJ/cm$^2$). The dramatic changes reflect the non-equilibrium phonon populations and their time dependence. a) Raw diffraction pattern of graphite along the [001] zone axis showing 6-fold symmetry of the Graphene planes. b) Differential scattering flat-field $\Delta I(\boldit{q},\tau)$ at a time before optical excitation indicating signal-to-noise. c) $\Delta I(\boldit{q}, 0.5~\textup{ps})$ provides a map of the relative strength of the $\boldit{q}$-dependent EPC coupling through the increased occupancy of strongly-coupled modes. Peaks in $\Delta I(\boldit{q}, 0.5~\textup{ps})$ at the $K-$points surrounding $\{2\bar{1}0\}$ (indicated) result from the increase in $K-A_1'$ population and outline the hexagonal BZ. 
Scattering from the $\Gamma-E_{2g}$ LO phonon is forbidden at $\{2\bar{1}0\}$, but strong coupling to the entire LO branch is evident in the vicinity of $\{200\}$ as ridges of intensity radiating from $\Gamma$ (the Bragg peak) to K points. d) - f) Non-equilibrium phonon dynamics: relaxation of the transient population of strongly-coupled modes. d) At 1.5~ps the peaks evident at $K-$points in panel c) have disappeared and diffuse intensity now appears halfway between $\{2\bar{1}0\}$ and the BZ edge (inset), but is still absent in the $M$ and $\Gamma_{\{2\bar{1}0\}}$ regions e) By 5~ps, the character of $\Delta I(\boldit{q},\tau)$ has changed dramatically to bands of intensity in the $\Gamma_{\{2 \bar{1} 0\}}-M-\Gamma_{\{200\}}$ direction approximately orthogonal to $\boldit{q}$, but the troughs near $\Gamma_{\{ 2 \bar{1} 0 \}}$ remaining. f) At 100~ps the $\Gamma_{\{ 2 \bar{1} 0 \}}-M-\Gamma_{\{200\}}$ bands have become sharper and the through at $\Gamma_{\{ 2 \bar{1} 0 \}}$ have filled in. Strong halos of diffuse intensity are present around the $\{100\}$ and $\{110\}$ families of peaks are evident (inset). These halos are weak, but present at 5~ps (inset, panel e)}
\label{fig:maps}
\end{figure*}

The photoinduced changes to the ultrafast electron scattering pattern, $\Delta I(\boldit{q},\tau) = \frac{I(\boldit{q},\tau)-I(\boldit{q},-\infty)}{I(\boldit{q},-\infty)}$, are shown for several points in time following photo-excitation in Fig.~\ref{fig:maps}.  The evolution of these patterns from 0.5 to 100 ps is striking and encodes detailed information on changes in the phonon system.  The diffuse scattering intensity at $\boldit{q}$ is modulated according to population dynamics of phonon modes with momentum $\boldit{k}$~\cite{Holt1999DeterminationSilicon}:
%
\begin{equation}
I(\boldit{q}) \propto 
	\sum_{j=1}\frac{n_{j,\boldit{k}}}{\omega_{j,\boldit{k}}}
    \underbrace{
    	\left|\sum_{s}\frac{f_s}{\sqrt{\mu_s}}\exp(-M_s)(\boldit{q}\cdot  \hat{\textbf{e}}_{j,s,\boldit{k}})\right|^2
    }_{\left|F_j(\boldit{q})\right|^2}.
\label{eqn:diffuseInt}
\end{equation}
%
The first sum is taken over $j$ phonon branches and the second is taken over $s$ atoms in the unit cell. $f_s$ is the atomic scattering factor, $\mu_s$ is the atomic mass, and $M_s$ is the Debye-Waller factor. Also, $\omega_{j,\boldit{k}}$ and $\hat{\textbf{e}}_{j,\boldit{k}}$ are the momentum-dependent phonon frequency and polarization for branch $j$. Finally, $n_{j,\boldit{k}}$ is the population of the phonon mode with frequency $\omega_{j,\boldit{k}}$. The UEDS patterns (Fig.~\ref{fig:maps}) provide an ultrafast snapshot of the change in $I(\boldit{q})$ for all $\boldit{q}$ proportional to the instantaneous occupancy of the $\omega_{j,\boldit{k}}$ modes, multiplied by the norm of the \textit{one-phonon structure factor}, $F_j(\boldit{q})$. In general, the occupancy of a specific branch, $n_{j,\boldit{k}}$, is not directly available from the inelastic scattering signal at $\boldit{q}$, since all modes $j$ contribute to $I(\boldit{q})$; diffuse scattering is momentum resolved but energy-integrated. However, the complete phonon band structure of Si has been determined by combining modeling and thermal diffuse x-ray scattering data~\cite{Holt1999DeterminationSilicon}. In addition, one can incorporate symmetry-imposed inelastic scattering selection rules~\cite{PerezMato1998,Kirov2003NEUTRONExperiments} which describe extinctions in $F_j(\boldit{q})$ at particular points of the diffuse scattering pattern for a given phonon mode. The extinctions depend on $\boldit{q}$ and the symmetry of the reduced wavevector $\boldit{k}$. If the symmetry group of the scattering vector is strict, several of the phonon modes can be inactive ($F_j(\boldit{q}) = 0$), reducing the number of phonon branches that can contribute at that point. Thus, graphite and other high-symmetry 2D materials are excellent candidates for UEDS. These selection rules have been used previously in inelastic x-ray diffraction experiments~\cite{Maultzsch2004} to measure the energy dependence of individual branches in the phonon dispersion of graphite. Here, we use the same selection rules to separate  the population dynamics of individual phonon branches without energy resolution.

\begin{figure}[t]
\includegraphics[width =\columnwidth]{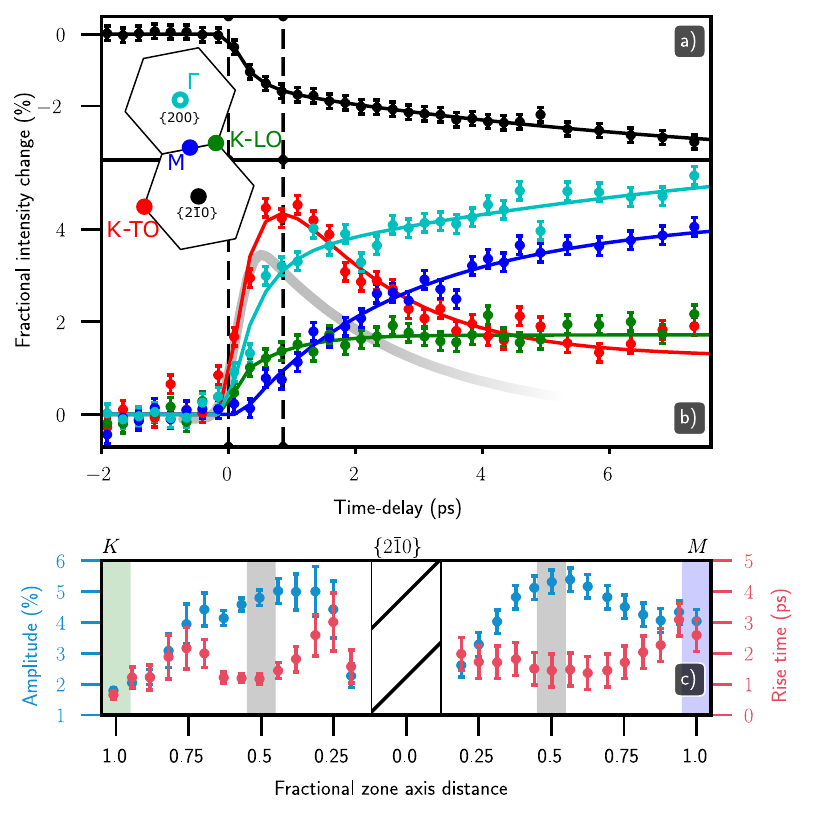}
\caption{Ultrafast electron diffuse scattering a) Intensity of the $\{2\bar{1}0\}$ Bragg peak showing profoundly non-exponential Debye-Waller dynamics.~\cite{Chatelain2014a} b) $\Delta I(\boldit{q},\tau)$ at select points in the BZ (inset). The rate of increase in the the population (EPC) of the TO $K-A_1'$ phonon (red, 280 fs) is faster than that for the LO $\Gamma-E_{2g}$ (cyan, 430 fs) and matches the fast Bragg peak dynamics. The population $K-$LO phonons (green, 730 fs) rises much slower that both TO $K-A_1'$ and LO-$\Gamma-E_{2g}$ phonons.  The rise in diffuse intensity at the M-point (blue, 2.1 ps) is almost an order of magnitude slower than that associated with the TO-$K-A_1'$ phonon.  The slow timescale decay evident in the Bragg peak~\cite{Chatelain2014a} and reported in earlier ARPES measurements~\cite{Tanaka2013} does not emerge from the dynamics of any single mode, but is a composite of the decay in population of the strongly coupled optical modes (e.g. red, 1.7 ps) and the increase in population of all other modes.  c) Diffuse intensity dynamics for points along the $K-\Gamma$ line, and $\Gamma-M$ line (inset). Rise time (blue) and amplitude (red) from single exponential fits to the early time dynamics are shown. Error bars represent covariance in fit parameters.  Points close to $\Gamma$ are not shown due to the interference of Bragg peak Debye-Waller dynamics of a).}\label{combofig}
\end{figure}

An attractive feature of UEDS is that a discrete, strongly-coupled mode yields a peak in the differential scattering maps at the associated BZ momentum position of that phonon at short delay times; electronic excitation energy initially flows preferentially to modes with strong EPC and are the first to show an increase in diffuse scattering (Fig.~\ref{fig:maps} c).  Here, diffuse scattering peaks ($\textup{FWHM} = 0.12$~\r{A}$^{-1}$) appear at the $K-$points along the reflection axes, near the $\{2\bar{1}0\}$ family of peaks (where scattering from the $K-A_1'$ mode is allowed) and along star-like ridges joining $\Gamma_{\{200\}}-K$ (where LO branch scattering is allowed). The phonon dispersion relation of graphite (Fig.~\ref{fig:introfig}~b) shows strong softening of the LO branch near $\Gamma$ and the TO branch near $K$ due to Kohn anomalies~\cite{Piscanec2004,Politano2015}~(Fig.~\ref{fig:introfig}). Earlier work suggested these strongly-coupled modes as the initial reservoir into which the electronic excitation energy flows~\cite{Kampfrath2005} and our results confirm that hypothesis. Time-resolved Raman has previously been employed to follow the the occupancy of the zone center $\Gamma-E_{2g}$ mode showing that it is indeed strongly-coupled~\cite{Yan2009,Yang2017}. Evidence for strong coupling to the off-zone-center $K-A_1'$ mode has previously only been indirect and the peaks in Fig.~\ref{fig:maps}~c) are the first direct observation.  In addition, Fig.~\ref{fig:maps}~c) indicates that coupling is strong for the entire LO branch between $\Gamma-K$ not only for $\Gamma-E_{2g}$ mode. 
The character of the differential diffuse scattering pattern changes dramatically through Fig.~\ref{fig:maps}~c-f) as the non-equilibrium phonon distribution evolves, demonstrating their profound sensitivity to the details of the phonon occupancies. The complete time-dependence of the diffuse intensity at selected points is shown in Fig.~\ref{combofig}. Scattering from the $K-A_1'$ mode is forbidden by symmetry at the $K-$points immediately proximate to the reflection axes (indicated by a green in the legend); only LO phonon scattering is observed at these points~\cite{Maultzsch2004}. Thus, $\Delta I(\boldit{q},\tau)$ at this point shows a qualitatively distinct time-dependence (Fig.~\ref{combofig}~b), green) versus $K-$points along the reflection axes at which scattering from the $K-A_1'$ mode is allowed (Fig.~\ref{combofig}~b), red).  This includes a much slower initial rise; 730 fs ($K-$LO) compared to 280 fs ($K-$TO).  Intensity near $\{200\}$ (Fig.~\ref{combofig}~b), cyan) reports on the occupancy of the strongly-coupled $\Gamma-E_{2g}$ LO mode at early times, and exhibits a slower rise (430 fs) than the $K-A_1'$ mode. For comparison the $\Gamma-E_{2g}$ phonon population determined using TR-Raman~\cite{Yan2009} is shown in grey. The Raman curve is nearly identical to that shown for the $K-A_1'$ phonon in terms of rise time and recovery, but differs from the data shown in cyan.  The slower rise time observed here is likely due to the higher excitation conditions used (12~mJ/cm$^2$ compared to 0.2~mJ/cm$^2$) which is known to weaken the Kohn anomaly and EPC at the $\Gamma$ point~\cite{Ishioka2008,Yan2009}. The different behavior of the cyan curve on picosecond time-scales is due to scattering from the low-frequency LA and TA phonon modes involved in the dominant decay channel of the $K-A_1'$ phonon (Fig.~\ref{fig:introfig}~b) and discussed further below. These LA/TA modes are not seen in the TR-Raman study, nor do they overlap with the signals measured at the $K-$point. 

We can estimate the effective temperature of the $K-$point TO mode, $T_{TO,K}$, using the measured increase in diffuse intensity and applying Bose-Einstein statistics to the mode population, $n_{j,\boldit{k}} = \cosh\left(\hbar\omega_{j,\boldit{k}}/2k_BT_{j,\boldit{k}}\right)$. We determine that the effective temperature of the $K-$point TO mode is $\sim1300~$K by 1 ps and also that by 10 ps it has cooled back down to $\sim450~$K. The strongly-coupled TO and LO modes reach a pre-equilibrium with the laser-generated carriers in $<1~$ps, while all other phonon modes remain at or near room temperature.

The diffuse scattering patterns in Fig.~\ref{fig:maps}~d-f) reveal the decay channels for the generated population of strongly-coupled optical phonons as they relax. The time-scales separation between the EPC into $K-A_1'$ and $\Gamma-E_{2g}$ ($200-400$ fs) and the subsequent decay out of these modes ($1-3$ ps) means that the diffuse scattering pattern at 1.5 ps maps their momentum-dependent decay probability in a manner analogous to the way in which the 0.5 ps pattern indicates the relative EPC strength.  The probability for the $\Gamma-E_{2g}$ phonon to decay to two phonons of momentum $\boldit{k}$ and $-\boldit{k}$ was previously computed using density functional perturbation theory~\cite{Bonini2007} for the distinct momentum and energy conserving channels indicated with arrows in Fig.~\ref{fig:introfig}~b). The results of these calculations compare well with $\Delta I(\boldit{q},\tau=1.5~\textup{ps})$ in Fig.~\ref{fig:maps}~d) (inset). The dominant decay channels are presented in Fig.~\ref{fig:introfig}~b). The hexagonal distribution of diffuse intensity approximately halfway between $\Gamma$ and the BZ edges is associated with the LA-TA and LA-LA decay channels. Peaks in evident along the $\Gamma-K$ lines are also in agreement with the expected location of maximum decay rate. Diffuse intensity around the $K-$points are associated with the TA-TA decay channel. The time-constants associated with each of these channels can be determined directly from the time-dependence of $\Delta I(\boldit{q},\tau)$ at the associated $\boldit{k}$ position in the BZ, shown for selected positions in Fig.~\ref{combofig}~b). The variation in the rise time and amplitude of $\Delta I(\boldit{q},\tau)$ in the BZ surrounding the $\{2\bar{1}0\}$ peak is given along the $\Gamma-K$ and $\Gamma-M$ directions in Fig.~\ref{combofig}~c). The dominant TA-LA and TA-TA decay channels in the $\boldit{k} = 0.5~K$ and $\boldit{k} = 0.5~M$ regions have the fastest time-constants ($1.0-1.5$ ps) and largest amplitude. 

\begin{figure}[t]
	\includegraphics[width =\columnwidth]{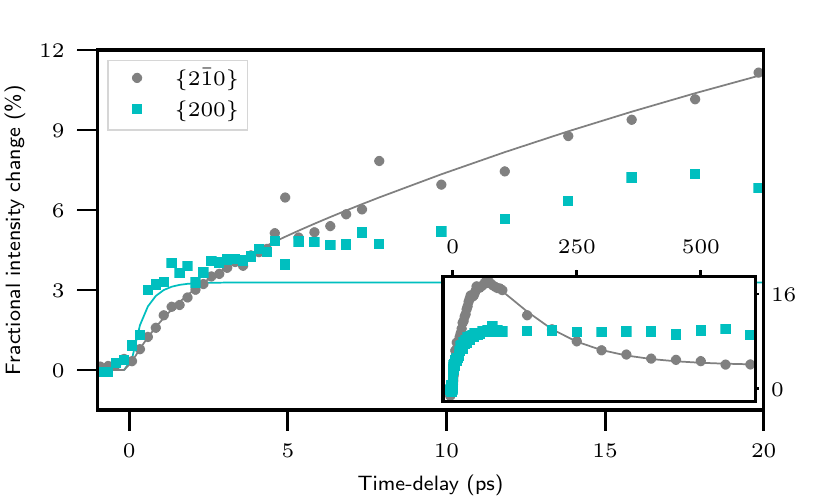}
    \caption{Diffuse intensity near $\Gamma_{\{ 2 \bar{1} 0\}}$ (grey) and $\Gamma_{\{ 200\}}$ (cyan). Early-time diffuse intensity changes demonstrate how time-resolution and selection rules can be used to separate the dynamics of phonon branches at the same momentum. The cyan curve displays a quasi bi-exponential character while the grey curve a single exponential rise (1.5~ps). The fast cyan component (430~fs, solid line) has been assigned to EPC to the $\Gamma-E_{2g}$ mode.  This signal is absent from the grey curve due to symmetry-imposed selection rules. The 1.5~ps rise in the grey curve (solid line) is associated with the low frequency LA/TA populations generated through $K-A_1'$ phonon decay.  The longer time constant in the cyan curve is a composite time-scale associated with the decay of the $\Gamma-E_{2g}$ population and the increase in the low frequency LA/TA populations. Inset: Diffuse intensity changes for the entire time range. The 34~ps rise in intensity in the grey curve is associated with the intraband thermalization of the non-equilibrium LA and TA populations generated through the decay of strongly-coupled $\Gamma-E_{2g}$ and $K-A_1'$ phonons.  The 115~ps decay most evident in the $\Gamma_{\{ 2 \bar{1} 0\}}$ data is associated with interband equilibration of the phonon populations in LA, TA and ZA bands. Solid grey line shows triexponential fit to the 1.5~ps rise, 34~ps rise, and 115~ps decay.}\label{fig:longtime-scale}
\end{figure}

The decay of the $K-A_1'$ phonon population is evident in the relaxation of the $K-$TO intensity shown in Fig.~\ref{combofig}~b) (red, 1.7 ps), is dominated by LO-LA and LA-TA channels. The LA-TA decay overlaps with that of the $\Gamma-E_{2g}$ phonon and is not easily separated, however, the LO-LA and LA-TA channels yield an increase in intensity adjacent to $\Gamma$ and $K$ points that can be readily identified.  Fig.~\ref{fig:longtime-scale} shows the time dependence of the diffuse intensity adjacent to the \{200\} and \{210\} peaks along the $\Gamma-K$ direction. Adjacent to \{210\}, scattering from $\Gamma-E_{2g}$ is forbidden by symmetry; thus, the fastest time-constant evident in the data, 1.5~ps (Fig.~\ref{fig:longtime-scale} (grey)), can only be assigned to the LO-LA and LA-TA decay channels of the $K-A_1'$ phonon. Adjacent to \{200\}, scattering from $\Gamma-E_{2g}$ is allowed and the early time dynamics appear approximately bi-exponential (Fig.~\ref{fig:longtime-scale}~inset (cyan)).  The observed fast time-constant behaviour is associated with the strong EPC to the $\Gamma-E_{2g}$ mode, as previously described.  The slower dynamics is assigned to a composite time-scale resulting from a decrease in intensity due to the decay of $\Gamma-E_{2g}$ phonons and an increase due to the LO-LA and LA-TA decay channels of the $K-A_1'$ phonon.  It is clear that time-resolution in UEDS can substitute for energy resolution when discriminating contributions from various phonon branches at a single momentum.

On longer time-scales the dynamics evident in the phonon system remain rich.  The early-time diffuse intensity bands oriented along $\Gamma-K$ directions  (Fig.~\ref{fig:maps}~c) and d)) evolve to become intensity bands oriented along $\Gamma-M$ directions. The two principle factors in these dynamics are that the decay of the strongly-coupled optical modes generates a profoundly non-equilibrium population of LA and TA phonons (Fig.~\ref{fig:introfig}~b), and that these branches are considerably softer along the $\Gamma-M$ direction than the $\Gamma-K$ in graphite. The non-equilibrium populations of LA and TA phonons thermalize within these bands by emission of lower frequency LA and TA phonons.  This relaxation channel is evident in the increase in the diffuse halos surrounding the Bragg peaks in Fig.~\ref{fig:maps}~e) and f) (inset), whose detailed time dependence is plotted in Fig.~\ref{fig:longtime-scale}. This relaxation of the phonon distribution in the acoustic branches occurs on a time-scale of 34~ps and is a phonon analog of the carrier thermalization that has previously been observed in TR-ARPES experiments~\cite{Gierz2015,Stange2015,Yang2017}.  
The lowest frequency branch of the in-plane phonon band structure consist of acoustic modes with a polarization normal to the graphene planes (ZA modes), to which the current experiments are insensitive due to electron beam orientation. Relaxation into the ZA branch and LA/TA interband relaxation is evident, however, as the decay of the diffuse halo intensity surrounding the $\{2\bar{1}0\}$ peak (Fig.~\ref{fig:longtime-scale}) which occurs on the 115~ps time-scale.  All of these time-scales are much faster than thermal transport of the laser deposited energy out of the probed volume, which occurs on the 10~\textmu s time-scale in the geometry of these experiments.

We have demonstrated that UEDS provides direct, momentum resolved measurements of the relative strength of EPC and PPC in graphite through the technique's ability to follow phonon population dynamics with femtosecond time resolution. Particularly notable is the unprecedented picture of phonon decay kinetics that the UEDS patterns provide, only a summary of which has been given here.  UEDS is also profoundly complementary to ultrafast ARPES. Together these methods can provide a complete picture of the dynamics within and between electron and phonon sub-systems, and help unravel the physics of complex phases where the intertwined nature the electron-lattice systems determine material properties.

\section{Acknowledgements}
This work was supported by the Natural Sciences and Engineering Research Council of Canada (NSERC), the Fonds de Recherche du Qu\'ebec - Nature et Technologies (FRQNT), the Canada Foundation for Innovation (CFI), and Canada Research Chairs (CRC) program.

\bibliographystyle{apsrev}
\bibliography{Mendeley}

\end{document}